\documentclass{ws-ijmpa}

\usepackage[super,compress]{cite}

\usepackage[T1]{fontenc} %pode retirar?
\usepackage{graphicx}

\usepackage{soul}
\usepackage{xcolor}
\usepackage{url}

\usepackage{multirow}
\usepackage{booktabs} % For better looking lines
\usepackage{threeparttable}

\usepackage{pdflscape}
\usepackage{hyperref}
\usepackage{sourcecodepro} % a monospaced font

\begin{document}
\markboth{C. de Oliveira \& V. de Souza}{Perspectives on multimessenger astrophysics}

%\catchline{}{}{}{}{}
%
%%%%%%%%%%%%%%%%%%%%%%%%%%%%%%%%%%%%%%%%%%%%%%%%%%%%%%%%%%%%%%%%%%%%

\title{Recent advances on multimessenger astrophysics: Centaurus A, GW 170817, and KM3-230213A}

\author{Cain\~a de Oliveira}
\address{Instituto de F\'isica de S\~ao Carlos, Universidade de S\~ao Paulo\\
S\~ao Carlos, S\~ao Paulo 13566-590, Brasil\\
olivcaina@gmail.com}
\author{Vitor de Souza}
\address{Instituto de F\'isica de S\~ao Carlos, Universidade de S\~ao Paulo\\
S\~ao Carlos, S\~ao Paulo 13566-590, Brasil\\
vitor@ifsc.usp.br}

\maketitle

%\begin{history}
%\received{Day Month Year}
%\revised{Day Month Year}
%\end{history}

\begin{abstract}
We review recent advances in multimessenger astrophysics, with particular emphasis on Centaurus A and on two pivotal events: the gravitational-wave detection GW~170817 and the ultra-high-energy neutrino KM3-230213A. Centaurus A is the prototype multimessenger source. The GW~170817 event, arising from a binary neutron star merger, marked a transformative moment in astronomy through the joint observation of gravitational waves and a broad spectrum of electromagnetic signals. The KM3-230213A neutrino, detected by the KM3NeT Collaboration, is the most energetic neutrino observed to date and poses significant challenges for current models, given its tension with null results from IceCube and the Pierre Auger Observatory. We assess astrophysical interpretations, including galactic, cosmogenic, and transient extragalactic sources, as well as implications for cosmic-ray acceleration. These cases underscore the scientific potential of high-energy multimessenger events in probing both the extreme universe and new physics.

\end{abstract}

\ccode{PACS numbers:}

%\tableofcontents

\section{Introduction}	

The development of robust multimessenger astrophysics offers a deeper understanding of the physical processes operating in the most extreme environments known in the Universe. Moreover, it holds great promise for the discovery of new physics, as it compels the application of physical theories at the limits of their validity. Two key aspects are particularly noteworthy: the combination of multiple messengers to investigate a given astrophysical source and its implications for general theoretical frameworks; and the study of correlations among different messengers to formulate hypotheses regarding their origin.

This review is structured around three case studies. The first, presented in Section~\ref{sec:cenA}, focuses on the radio galaxy Centaurus~A (Cen~A). This section compiles and compares models of ultra-high-energy cosmic ray (UHECR) acceleration in Cen~A, highlighting how the multimessenger context can enhance our understanding of particle acceleration mechanisms in this source, potentially extending to the ultra-high-energy regime.

Section~\ref{sec:GW170817} presents the GW~170817 event, the first multimessenger observation involving gravitational waves and electromagnetic signals. This transient event provided confirmation of long-standing astrophysical models, enabled tests of fundamental physics, and contributed to theoretical advancements, thereby serving as a paradigmatic example of the potential of multimessenger astrophysics.

Section~\ref{sec:KM3230213A} discusses the recent detection of the UHE neutrino KM3~230213A and its impact on the scientific community. Following its official announcement, the event prompted an intense scientific effort to interpret the observation and identify its possible origin.

\section{Centaurus~A: The multimessenger source}
\label{sec:cenA}

Centaurus A (Cen~A) is the closest radio galaxy to the Milky Way ($3.8 \pm 0.1$~Mpc)\cite{harris_cenA}, classified as a Fanaroff-Riley type I (FR I) galaxy\cite{israel1998centaurus}. As a jetted active galactic nucleus (AGN), multiple acceleration sites may contribute to the nonthermal cosmic ray (CR) spectrum. Particle acceleration in Cen~A has been both proposed and observed across various spatial scales and energy ranges\cite{croston2009_shock,hardcastle_lobes_09,Abdo_2010,matthews_backflow,hess2020resolving,zhang_murase_2023,fermi_innerlobes,fermi_2010_giant_lobes}. 

Its proximity enables electromagnetic observations over a broad range of spatial scales (from sub-parsec to megaparsec), revealing diverse bubble-shaped structures: two giant outer lobes, a Northern Middle Lobe, two Inner Lobes, in addition to the jet and core. These structures have been previously studied as potential CR accelerators, including to ultra-high energies. In principle, distinct or combined acceleration mechanisms may operate at different sites, with the total emission arising from a superposition of several processes. For instance, along the large-scale jet, CRs may be accelerated by multiple shocks, shearing flows, and turbulent regions, yielding an injection spectrum shaped by a combination of shear, and first- and second-order Fermi acceleration mechanisms\cite{Mbarek_2021,Seo_2023,Seo_2024}. The acceleration timescale and resulting source spectrum influence the spectra of the associated neutrinos and $\gamma$ rays produced via interactions during acceleration, escape from the site, and, at extreme energies, propagation through the extragalactic environment. In this context, the detection of multiple messengers can enhance our understanding of particle acceleration throughout the source.

As the nearest radio-loud AGN, Cen~A has long been studied via electromagnetic observations. \textit{Chandra} X-ray data\cite{croston2009_shock}, for instance, revealed shock-accelerated particles around the inner south-west radio lobe. The detection of $\gamma$ rays by \textit{Fermi}-LAT\cite{Abdo_2010} and HESS\cite{hess2020resolving} indicates the presence of electrons with Lorentz factors of $\sim10^8$ along the jet. Analyses of \textit{Fermi}-LAT data also suggest a possible association of the $\gamma$-ray signal with the jet and inner lobes\cite{fermi_innerlobes}. \textit{Fermi}-LAT measurements additionally revealed relativistic electrons ($\sim0.1$–$1$~TeV) in the Giant Lobes\cite{fermi_science_giant_lobes}. 

The identification of nonthermal electrons in a given region has been consistently accompanied by estimates of the energy limits for protons potentially accelerated at the same sites. In the inner south-west radio lobe\cite{croston2009_shock}, shock properties suggested an upper limit of $\sim2$~EeV for protons. Achieving extreme energies of $\sim100$~EeV would require an unreasonably large magnetic field amplification; however, the possibility of a heavier composition, as indicated by current data\cite{mass_deep_learning}, could mitigate the required amplification. In the core region, the maximum energy for protons was estimated to be $\lesssim 40$~EeV, increasing for heavier nuclei\cite{Abdo_2010}. In optimistic scenarios, protons accelerated in the giant lobes could reach $\sim100$~EeV\cite{hardcastle_lobes_09}.

For many years, Cen~A has been considered a candidate for accelerating CRs to the ultra-relativistic regime. Speculations about its role as a UHECR source gained prominence after the 2007 report of a correlation between the directions of nearby AGNs ($<75$~Mpc) and the arrival directions of UHECRs with energies above 60~EeV, as detected by the Pierre Auger Observatory\cite{auger_agn_2007,auger_agn_2008}. At the time, the proximity of several events in the direction of Cen~A led to estimates of the expected flux of UHECRs and neutral secondaries from the source\cite{cuoco_2008,Kachelrieß_2009,hardcastle_lobes_09}. Subsequent data revealed a decrease in the overall significance of the correlation between nearby AGNs and UHECR arrival directions\cite{ABREU2010314}. 

Nevertheless, a distinct excess of events with energies $>55$~EeV has been consistently observed in the direction of Cen~A — referred to as the ‘Centaurus region’. This excess currently has a post-trial significance of $3.1\sigma$ within a $27^\circ$ radius around $\left(\ell_{sg}, b_{sg} \right)=\left(163^\circ, -3^\circ \right)$\footnote{Coordinates given in supergalactic coordinates. For comparison, Cen~A lies at $\left(159.75^\circ, -5.25^\circ \right)$.} for events above 63~EeV, and persists for energies down to 20~EeV\cite{thepierreaugercollaboration2024fluxultrahighenergycosmicrays}. 

Although the origin of this hotspot remains unclear, its detection reinforces the plausibility of Cen~A as a UHECR accelerator, which, if confirmed, would constitute the first identified source of UHECRs\footnote{Alternative hypotheses include starburst galaxies in the Centaurus region or deflected flux from other sources due to the Galactic Magnetic Field\cite{deOliveira_2023,sombrero2024}}.

In addition to the UHECR hotspot, the detection of $\gamma$ rays has stimulated considerable research efforts. The energy flux measured by \textit{Fermi}-LAT shows an excess relative to a single-zone Synchrotron Self-Compton (SSC) model of the core, which otherwise fits the electromagnetic spectrum from radio frequencies up to $\sim$~GeV $\gamma$ rays. Multiple hypotheses have been proposed to explain this excess, though no consensus has been reached regarding a hadronic or leptonic origin (see reviews\cite{rieger2017gamma,tev_radiogal}). Many of these studies assume that the very-high-energy (VHE) radiation originates in the core region. However, recent observations by HESS\cite{hess2020resolving} detected a $\gamma$-ray signal best described by an elliptical morphology, consistent with an extended emission along the jet rather than a radially symmetric core origin. This morphological evidence suggests that particle acceleration along the jet is primarily responsible for the VHE $\gamma$-ray emission, thereby constraining the possible contribution from the core and, consequently, the related acceleration models.

In this section, we compare models for UHECR production in Cen~A from a multi-messenger perspective, focusing on scenarios with a hadronic origin for the $\gamma$-ray emission, thereby allowing for an association between hadronic CR acceleration and the observed $\gamma$-ray signal. Models positing the core as the primary acceleration region are excluded, given that the morphological characterization disfavors many such scenarios.

\begin{landscape}
\begin{figure}[p]
    \centerline{\includegraphics[width=1.5\textwidth]{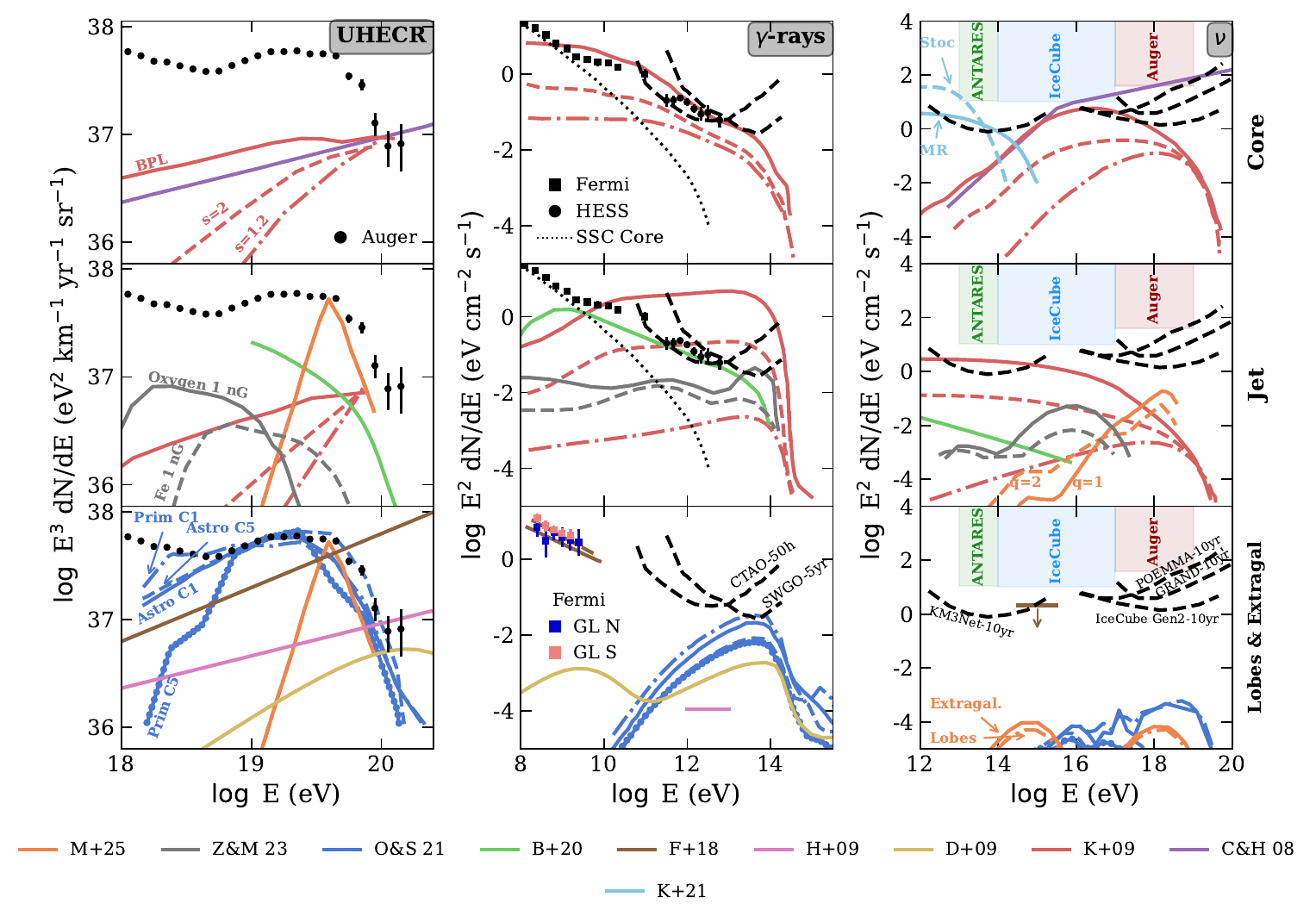}}
    \caption{Multimessenger signals for Cen~A. \textit{Left}: UHECR spectral models and data from the Pierre Auger Observatory\cite{Abreu_2021}. \textit{Center}: $\gamma$-ray models and observations from \textit{Fermi}-LAT and HESS\cite{hess_fermi}; sensitivity curves of CTAO\cite{ctao_link} and SWGO\cite{swgo} are also shown. \textit{Right}: Neutrino models and upper limits from ANTARES\cite{PhysRevD.96.082001}, IceCube\cite{Aartsen_2017}, and the Pierre Auger Observatory\cite{Aab_2019_point_neutrinos}; expected sensitivities from KM3NeT\cite{km3net_sensitivity}, POEMMA\cite{poemma}, GRAND\cite{grand}, and IceCube-Gen2\cite{icecubegen2_sensitivity} are also included following M+25. The rows correspond to different acceleration regions: core (\textit{top}), jet (\textit{middle}), and lobes (\textit{bottom}). Secondary production due to UHECR propagation is included in the lower panels. See the main text for detailed descriptions of the models and legend.}
    \label{fig:CenA}
\end{figure}
\end{landscape}

Figure~\ref{fig:CenA} presents flux predictions for various messengers based on several models proposed in the literature for UHECR acceleration in Cen~A. The rows correspond to different acceleration regions. The first column shows the predicted UHECR signal, while the second and third columns display the resulting VHE $\gamma$-ray and neutrino fluxes, respectively. Each color represents a different author, while distinct symbols indicate varying assumptions. A brief description of each model follows below.

\textbf{Cuoco \& Hannestad (2008) (C\&H 08)}\cite{cuoco_2008}: Investigates the connection between UHECRs and neutrino emission resulting from acceleration and interactions within the nucleus. The model is normalized to the two UHECR events ($>60$~EeV) detected by the Pierre Auger Observatory within approximately $3^\circ$ of Cen~A. The source spectrum is assumed to follow a power law $E^{-2.7}$, consistent with the Auger data available at the time. During acceleration, protons primarily interact with the nuclear radiation field, based on observational data. While protons are magnetically confined within the acceleration region, neutrons escape and contribute to the UHECR flux. The resulting neutrino flux is considered an upper limit under this framework.

\textbf{Kachelrieß et al. (2009) (K+09)}\cite{Kachelrieß_2009}: Explores acceleration both along the jet and near the core. Using parameterizations for local photon and matter densities, the study estimates neutrino and photon fluxes under the assumption that protons are accelerated up to $10^{20}$~eV. The model adopts the same normalization as C\&H08. Near the core, the ultraviolet radiation field dominates as a scattering target, while proton-proton interactions are more relevant along the jet. Three spectral scenarios are considered: (i) a non-relativistic shock model with $E^{-2}$ (labelled $s=2$ in Fig.~\ref{fig:CenA}); (ii) a broken power law, $E^{-2}$ below $E_b$ and $E^{-2.7}$ above, for $E_b=10^{17}$~eV and $E_b=10^{18}$~eV (labelled BPL in Fig.~\ref{fig:CenA}); and (iii) a flat spectrum $E^{-1.2}$, consistent with linear acceleration ($s=1.2$ in Fig.~\ref{fig:CenA}). For the broken power-law model, the resulting photon flux was already disfavored based on \textit{Fermi}-LAT upper limits at the time. The study concluded that neutrino detection was marginally plausible, but photon signals should be observed first—a prediction later confirmed by \textit{Fermi}-LAT and HESS. The analysis also emphasized the difficulties introduced by uncertainties in the UHECR event count, composition (particularly heavy nuclei), the assumed $E_{\rm max}=10^{20}$~eV, and magnetic deflections.

\textbf{Hardcastle et al. (2009) (H+09)}\cite{hardcastle_lobes_09}: Considers UHECR acceleration within the giant lobes of Cen~A and estimates the resulting $\gamma$-ray flux. Proton-proton interactions within the lobes are identified as the dominant mechanism for producing hadronic $\gamma$ rays. The lack of precise measurements of the thermal proton number density introduces significant uncertainty. Assuming $n \sim 10^{-4}~\rm{cm}^{-3}$ and a UHECR spectral index $s=2$, a conservative lower bound of $\sim 10^{-17}$~erg~cm$^{-2}$~s$^{-1}$ at $10$~TeV is set for the $\gamma$-ray flux. Using the C\&H08 UHECR spectrum, the predicted $\gamma$-ray flux from the lobes reaches $\sim 3 \times 10^{-16}$~erg~cm$^{-2}$~s$^{-1}$ (as shown in Fig.~\ref{fig:CenA}).

\textbf{Dermer et al. (2009) (D+09)}\cite{Dermer_2009}: Evaluates the high-energy cascade photon flux arising from UHECRs propagating toward Earth. The model is normalized using the two UHECR events ($>60$~EeV) detected near Cen~A, assuming a proton composition. The UHECR spectrum follows $E^{-2.2}$ with an exponential cutoff at $200$~EeV. Interactions with the cosmic microwave background (CMB), extragalactic background light (EBL), and a $1$~nG intergalactic magnetic field are included.

\textbf{Fraija et al. (2018) (F+18)}\cite{fraija2018}: Argues that the Giant Outer Lobes constitute the most favorable site for accelerating cosmic rays to UHEs. Protons and helium nuclei are found insufficient to meet the luminosity constraints, suggesting heavier nuclei, such as carbon, as more plausible candidates. The UHECR flux is normalized to 14 events ($>58$~EeV) within a $15^\circ$ radius around Cen~A, based on Auger data from 2015\cite{Aab_2015}. The CR spectrum is extrapolated to lower energies, and the secondary $\gamma$-ray flux produced within the lobes is computed. The potential to explain the IceCube event IC35 is explored, but the resulting neutrino flux remains too low. Hadronic interactions are found to be more efficient than photohadronic processes for neutrino production in this scenario.

\textbf{Banik et al. (2020) (B+20)}\cite{banik_cenA_2020}: Aims to jointly explain the observed VHE $\gamma$-ray emission and predict the corresponding UHECR flux from Cen~A. The authors employ a \textit{proton blazar-inspired model}, in which electrons, protons, and nuclei are accelerated in the AGN jet. Motivated by the coincidence of a UHECR excess in the direction of Cen~A and a spectral hardening in the $\gamma$-ray spectrum, the model assumes a hadronic origin for the VHE component. Neutral secondaries are generated by proton–proton interactions within the jet. The $\gamma$-ray emission from a moving spherical blob in the kiloparsec-scale jet is combined with the leptonic emission from a sub-parsec blob to reproduce the full spectral energy distribution (SED) of Cen~A. Only the hadronic contribution is shown in Fig.~\ref{fig:CenA}. The UHECR mass composition is fitted to $\langle \ln A \rangle$ values reported by the Pierre Auger Collaboration.

\textbf{de Oliveira \& de Souza (2021) (O\&S 21)}\cite{de2021probing}: Employ a phenomenological UHECR spectrum based on jet power estimations to calculate upper limits for the associated neutrino and $\gamma$-ray fluxes produced during extragalactic propagation. Several chemical compositions (labelled C1–C5) and extragalactic magnetic field configurations (AstroR, Prim2R, and Prim) are considered to explore the model's parameter space.

\textbf{Zhang \& Murase (2023) (Z\&M 23)}\cite{zhang_murase_2023}: Propose that the VHE $\gamma$ rays detected by Fermi-LAT and HESS originate from UHECRs accelerated along the jet, which subsequently interact with ambient radiation fields. Leptohadronic processes—photodisintegration, photomeson production, Bethe-Heitler pair production, and electromagnetic cascades—are modeled by solving coupled transport equations that include all relevant species: photons, electrons, neutrinos, neutrons, protons, and nuclei. The authors suggest that de-excitation $\gamma$ rays resulting from photodisintegration offer promising detection prospects for CTAO and SWGO\footnote{Although LHAASO is also considered, Cen~A lies outside its field of view.}. The model shows reasonable agreement with \textit{Fermi}-LAT and HESS data when assuming the injection of either oxygen or iron nuclei. However, the required UHECR luminosity is one to two orders of magnitude above current estimates of Cen~A's jet power ($\sim10^{43}$~erg/s), though still below the Eddington luminosity ($\sim7 \times 10^{45}$~erg/s) for the central SMBH.

\textbf{Mbarek et al. (2025) (M+25)}\cite{mbarek_cenA_2025}: Estimate the neutrino flux from Cen~A using a bottom-up approach. A 3D magnetohydrodynamical model of a radio galaxy jet is developed, incorporating standard parameters such as density contrast, Mach number, Alfvénic Mach number, magnetization, and effective Lorentz factor. An inner spine region is assumed with an effective Lorentz factor of $\sim3.5$ and a magnetic field strength of $\sim100~\mu$G. The resulting UHECR spectrum is normalized to represent different levels of contribution to the total UHECR flux ($3\%$ to $100\%$). Neutrino production is calculated based on interactions within the jet and cocoon, residence in the outer lobes, and extragalactic propagation, assuming a CNO-like composition. Neutrino production via espresso and shear acceleration is sensitive to the injection slope of escaping UHECRs, and results are shown for $q = 1,2$. Overall, the flux remains below the detection thresholds of current and upcoming observatories, except in the most extreme scenario, where IceCube-Gen2 might detect the signal after a decade of observation.

\vspace{0.25cm}

Originally, the models C\&H08\cite{cuoco_2008}, H+09\cite{hardcastle_lobes_09}, K+09\cite{Kachelrieß_2009}, D+09\cite{Dermer_2009}, F+18\cite{fraija2018}, and B+20\cite{banik_cenA_2020} normalized the UHECR flux based on the number of events detected in the direction of Cen~A by the Pierre Auger Observatory. In Figure~\ref{fig:CenA}, we re-normalized the flux using updated observational data from the Pierre Auger Collaboration. The exposure of the observatory has increased substantially, from approximately $9,000$~km$^2$~yr~sr in 2007 (used in C\&H08 and K+09) to $135,000$~km$^2$~yr~sr in the 2022 dataset\cite{thepierreaugercollaboration2024fluxultrahighenergycosmicrays}. We adopt the excess of events within the Cen~A hotspot region—characterized by a post-trial significance of $3.1\sigma$ — as a unified normalization reference. The energy threshold for normalization is selected based on the specific energy range most appropriate for each model. Table~\ref{tab:norm} presents a comparison of the original and updated normalization parameters. Models O\&S\cite{de2021probing} and M+25\cite{mbarek_cenA_2025}, which correspond to upper-limit estimates, remain unchanged.

We choose to normalize the models using the number of events observed in the Cen~A hotspot. However, significant uncertainties remain regarding the expected deflections caused by the extragalactic magnetic field (EGMF). These arise both from the poorly constrained EGMF intensity and structure and from the unknown composition of UHECRs. Despite these uncertainties, the hypothesis that Cen~A could be the dominant source of UHECRs remains plausible (e.g.,~\citen{silvia_cenA_2024}).

\begin{table}[t]
\centering
\tbl{Data used to normalize different models of UHECR in Cen~A.}
{\begin{tabular}{@{}ccccccc@{}} \toprule
%Model &  & Old data & &  & Current data & \\
%& Auger data & Energy range (EeV) & $\langle N \rangle$ & Auger data & Energy range (EeV) & $\langle N \rangle$ \\ \colrule
\multirow{2}{*}{Model} & \multicolumn{3}{c}{Old data} & \multicolumn{3}{c}{Current data} \\ \cmidrule(lr){2-4} \cmidrule(lr){5-7}
& Data\tnote{a} & Energy range (EeV)\tnote{b} & $\langle N \rangle$\tnote{c} & Data & Energy range (EeV) & $\langle N \rangle$ \\ \colrule
C\&H08, H+09, K+09, D+09 & $2007$ & $>60$ & 2 & $2022$ & $>63$ & 19.7\\
F+18 & $2014$ & $>58$ & 14 & $2022$ & $>63$ & 19.7\\
B+20 & $2009$ & $55-85$ & 9.8 & $2022$ & $>50$ & 38.6\\ \botrule
\end{tabular}} \label{tab:norm}
\end{table}

Some general conclusions can be drawn. The contribution of Cen~A to the observed UHECR flux remains uncertain from both theoretical and phenomenological perspectives. Models C\&H08, K+09, and M+25 predict a significant contribution only at the highest energies ($\sim10^{19.5}$~eV). This behavior is partly attributable to the absence of exponential cutoffs and propagation effects in the spectra presented by C\&H08 and K+09. In contrast, B+20 predicts a more substantial contribution at lower energies. Since O\&S21 derive a best-fit to the observed UHECR spectrum, their model yields the highest Cen~A contribution overall. It is worth emphasizing that many models assume a source injection spectrum, which is typically modified during escape from the source and subsequent extragalactic propagation.

The associated VHE $\gamma$ rays serve as critical messengers for elucidating particle acceleration processes in Cen~A and offer promising prospects for detection by future observatories such as CTAO and SWGO. Several scenarios proposed by K+09 are already disfavored based on $\gamma$-ray data from \textit{Fermi}-LAT and HESS. In particular, HESS morphological observations suggest that the emission originates along the jet rather than from the core. Consequently, the broken power-law spectrum accelerated in the core—which accounts for a significant fraction of the \textit{Fermi}-LAT and HESS fluxes—can be excluded. This model also yields one of the highest predicted neutrino fluxes among all those shown in Figure~\ref{fig:CenA}. If UHECR acceleration occurs in the core under this model, the UHECR flux must be significantly reduced, thereby severely constraining the expected neutrino flux and rendering detection by future observatories improbable. In fact, this broken power-law core model was already constrained at the time of its original publication\cite{Kachelrieß_2009}. Similar issues arise for the jet-acceleration models employing a broken power law or a simple power law with index $s=2$, both of which overproduce the $\gamma$-ray flux relative to HESS and \textit{Fermi}-LAT measurements. In the case of the broken power-law model, the predicted UHECR flux must be reduced by a factor of $\sim10^3$ to comply with the $\gamma$-ray constraints, which severely limits the accompanying UHECR and neutrino signals from the K+09 scenarios.

Overall, the results suggest that the observed VHE $\gamma$-ray emission is likely associated with UHECR acceleration. Model B+20 achieves good agreement with both the $\gamma$-ray and neutrino data, while M+25 does not provide predictions for the VHE $\gamma$-ray flux. Model O\&S21 indicates that $\gamma$ rays produced during UHECR propagation from Cen~A may become detectable by CTAO and SWGO. The predicted signal is primarily dependent on the assumed source composition, and to a lesser extent on the extragalactic magnetic field. This opens a compelling avenue to use $\gamma$ rays above $\sim10$~TeV to constrain the contribution and composition of a potential UHECR flux originating from Cen~A.

The expectations for detecting neutrinos associated with UHECRs accelerated in and propagating from Cen~A are generally low. Model C\&H08 predicts the highest neutrino flux, which is likely already excluded by point-source limits from IceCube and the Pierre Auger Observatory. Indeed, the model adopts an overly optimistic scenario in which the composition is purely protons resulting from neutron decay, escaping the source. Models K+09 are similarly constrained by $\gamma$-ray data, positioning M+25 as the most promising scenario for future neutrino detection.

Although neutrinos with energies exceeding 10~PeV are unlikely to be observed, TeV neutrinos may serve as more viable messengers of particle acceleration in the inner regions of the source. One of the most promising predictions for a neutrino signal from Cen~A arises from the Magnetized Coronae model, proposed by Kheirandish et al. (2021) (K+21)\cite{Kheirandish_2021}. This model suggests that protons can be accelerated in the magnetized coronae of AGNs, leading to a multi-messenger signature testable in nearby Seyfert galaxies such as Cen~A. In this framework, a magnetized corona, situated above the accretion disk in the standard disk–corona structure of AGNs, facilitates proton acceleration via stochastic or magnetic reconnection processes. The maximum proton energy is constrained by cooling, advection, and diffusive escape, reaching up to the PeV scale. Proton-proton and proton-photon interactions yield secondary neutrinos and $\gamma$ rays; however, the region is expected to be optically thick to $\gamma$ rays, making the neutrino flux predominantly determined by the intrinsic X-ray luminosity.

The K+21 model includes free parameters calibrated to reproduce the neutrino signal associated with the Seyfert 2 galaxy NGC~1068\cite{ngc1068_icecube}. Of the three scenarios considered, two are displayed in Figure~\ref{fig:CenA}: one based on stochastic acceleration with high cosmic-ray pressure in the corona ("Stoc"), and the other based on magnetic reconnection acceleration ("MR"). The model is generalized to other bright X-ray Seyfert galaxies, selected from the BAT AGN Spectroscopic Survey, with the intrinsic X-ray luminosity used to compute the expected neutrino flux. Within this sample, Cen~A is the fourth-brightest source and the most promising candidate in the Southern sky, with predictions indicating detectability by the KM3NeT observatory within one to three years of observation.

Another viable scenario involves neutrino production from protons accelerated in radiatively inefficient accretion flows (RIAFs) of low-luminosity AGNs (LLAGNs)\cite{Kimura_2015}. Similar to the Magnetized Coronae model, this scenario does not aim to explain the origin of UHECRs. RIAFs occur when the mass accretion rate onto the supermassive black hole is low ($1$--$10\%$) relative to the Eddington rate. In this environment, protons undergo stochastic acceleration by turbulent magnetic fields, with the maximum energy limited by escape, yielding a characteristic energy of $E_{\rm p,~eq}\sim10$~PeV. Proton–proton interactions subsequently produce secondary neutrinos and $\gamma$ rays. This model was applied to Cen~A by Fujita et al.\cite{fujita_riaf_2015}, who modeled $\gamma$-ray production from interactions within the central molecular zone (CMZ) surrounding the supermassive black hole. The resulting $\gamma$-ray spectrum was found to agree reasonably well with HESS data above $\sim1$~TeV. In this scenario, $E_{\rm p,~eq}$ reaches approximately $7.9$~TeV, with a correspondingly faint neutrino flux.

\begin{figure}
    \centerline{\includegraphics[width=0.6\textwidth]{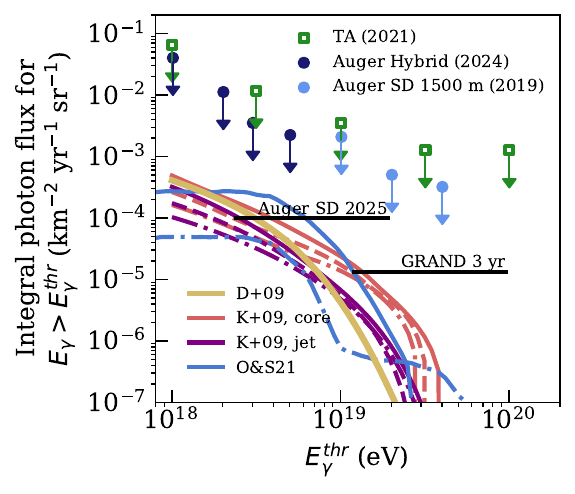}}
    \caption{Integral photon fluxes from various models compared to current upper limits from the Telescope Array (TA\cite{TA_uhe_phot}) and the Pierre Auger Observatory\cite{auger_uhe_phot} (Auger Hybrid and Auger SD 1500~m). Projected ideal upper limits from future observations by Auger and GRAND\cite{grand} are also shown. Line styles for models K+09 and O\&S21 correspond to those used in Fig.~\ref{fig:CenA}.}
    \label{fig:CenA_integral_phot}
\end{figure}

In addition to the messengers shown in Fig.~\ref{fig:CenA}, undetected ultra-high-energy (UHE) $\gamma$ rays can serve as valuable probes for constraining models and understanding particle acceleration in astrophysical sources. For Cen~A, the integral photon flux predicted by the various models is compared with existing upper limits in Fig.~\ref{fig:CenA_integral_phot}. As illustrated, current experimental sensitivities are insufficient to constrain the majority of the models. However, the anticipated upper limits from the Pierre Auger Observatory by 2025 and from three years of GRAND operation are expected to significantly enhance the observational capabilities. These improvements may enable the first detection of UHE photons from Cen~A or impose stringent constraints on the underlying particle acceleration scenarios.

\section{Gravitational Waves -- GW~170817 -- The Long-Awaited Messenger}
\label{sec:GW170817}

The detection of GW~170817\cite{GW170817_virgoligo}, followed by an extensive electromagnetic (EM) observation campaign\cite{GW_manycollab}, firmly established gravitational waves (GWs) as a viable and essential signal in the context of multi-messenger (MM) astronomy. The GW signal was detected on August 17, 2017, by the Advanced LIGO and Advanced Virgo detectors and was attributed to the inspiral and coalescence of a binary neutron star (BNS) system. The event was associated with the short gamma-ray burst GRB~170817A, detected $1.75$~s later by \textit{Fermi}-GBM and INTEGRAL, followed by a radio to X-ray afterglow and the kilonova AT~2017gfo, observed across ultraviolet (UV), optical, and near-infrared (NIR) wavelengths. The identification of the optical counterpart enabled the determination of the host galaxy and its distance, estimated at $40^{+8}_{-14}$~Mpc. This event inaugurated a new chapter in MM astrophysics and reached unprecedented significance. The scientific impact is evident: at least 1,446 publications in the \texttt{arXiv} repository cite GW~170817, with 397 mentioning it in the title\footnote{Considering the period between September 2017 and May 2025}—an average of approximately four papers per month.

The combined detection of GW and EM signals, along with the extensive multi-wavelength (MW) campaign, significantly advanced the understanding of short gamma-ray bursts (sGRBs) (see, e.g., Refs.~\citen{ciolfi_GW,lazzati_GW,radice_GW} for reviews). The event demonstrated that BNS mergers produce sGRBs and confirmed their role in r-process nucleosynthesis of heavy elements ($A > 140$) in a radioactively powered kilonova transient\cite{metzger2020kilonovae}. Beyond the astrophysical insights, the event allowed for novel constraints on fundamental physics, including the equation of state for ultra-dense matter\cite{raithel_GW}, dark energy\cite{GW_dark}, and modified gravity theories\cite{GW_mond,GW_mond2}.

In this section, we briefly summarize GW~170817 and its associated kilonova AT~2017gfo, and highlight how the combination of GW and multi-wavelength EM signals enabled the construction of a rich physical picture. The following summary presents the most widely accepted interpretation. For detailed discussions, limitations, and alternative views, the reader is referred to reviews such as Refs.~\citen{GW_manycollab,metzger2020kilonovae,ciolfi_GW,lazzati_GW,radice_GW,Branchesi2018}. The merger and prompt $\gamma$-ray emission are used to guide the temporal sequence of the event.

\subsection{Inspiral Phase: Pre-prompt Emission / Pre-merger — GW}

The characteristic GW signal of GW~170817, exhibiting increasing frequency and amplitude, revealed a binary compact object merger. Such systems possess a dynamic spacetime geometry that induces energy loss through gravitational wave emission. The inferred component masses are consistent with a binary neutron star (BNS) system; however, the GW data alone do not unambiguously distinguish between neutron star–neutron star (NS–NS), neutron star–black hole (NS–BH), or black hole–black hole (BH–BH) mergers. Nevertheless, the rich EM signal observed following the merger strongly indicates the presence of at least one neutron star, most likely confirming a BNS merger\cite{radice_GW,lazzati_GW,ciolfi_GW}.

\subsection{Remnant: Pre-prompt Emission / Post-merger — EM}

The merger produced a compact remnant whose nature remains uncertain, though it is most likely a meta-stable hypermassive neutron star (NS) that subsequently collapsed into a black hole (BH) within $\lesssim 0.1$–$1$~s. A prompt collapse into a BH, which is typically electromagnetically silent, is disfavored by the substantial kilonova ejecta inferred from observations. Conversely, the detection of a lanthanide-rich disk wind is in tension with the presence of a long-lived remnant. The energetics associated with the GRB and kilonova are inconsistent with the formation of a supermassive or stable NS, while the absence of an X-ray excess during the afterglow phase argues against the formation of a magnetar. The relativistic jet implied by the GRB detection at short time scales supports the existence of a spinning BH surrounded by a massive accretion disk. Although the post-merger remnant emits GWs, the expected frequencies lie outside the sensitivity range of the detectors available at the time, precluding direct detection. While no GW signal was used to draw the above conclusions, future GW observations will be essential for definitively determining the remnant's nature\cite{lazzati_GW,metzger2020kilonovae}.

\subsection{Wind and Jet Ejecta: Pre-prompt Phase — EM}

The merger is expected to produce both wind and jet ejecta. The jet is believed to have formed within $\lesssim1$~s after the merger, with an estimated viewing angle of $\sim15^\circ$–$35^\circ$ relative to our line of sight. The wind component likely preceded the jet and may have been partially emitted before the merger. As the jet propagates through the wind, it inflates a cocoon, resulting in a structured outflow. The jet, wind, and cocoon structure are all essential ingredients in modeling the subsequent EM observations\cite{lazzati_GW}.

\subsection{GRB~170817A: Prompt Emission — $\gamma$ rays}

A $\gamma$-ray pulse, characteristic of a GRB, was detected $1.74 \pm 0.5$~s after the GW signal by \textit{Fermi}-LAT and INTEGRAL. This initial pulse was followed by broader emission lasting approximately 2~s. The $\gamma$-ray luminosity was found to be $\sim10^4$–$10^5$ times fainter than that of typical sGRBs. The prevailing interpretation attributes this to an off-axis jet, with the observed emission originating from the cocoon rather than the jet core. A delay in $\gamma$-ray emission is expected: after the jet forms, it must propagate through the surrounding wind, reach an optically thin region, and dissipate energy as EM radiation. The time delay between the GW and $\gamma$-ray signals has been used to constrain alternative theories of gravity\cite{lazzati_GW,Branchesi2018}.

\subsection{Kilonova: Post-prompt Emission — UV/Optical/NIR}

The kilonova is characterized by a quasi-isotropic and quasi-thermal emission powered by the radioactive decay of heavy elements synthesized in neutron-rich material ejected during the merger. Approximately 11 hours post-merger, a rapidly fading thermal spectrum peaking in the UV was detected across UV and optical wavelengths — commonly referred to as the \textit{blue kilonova}. Several days later, the emission was dominated by a thermal peak in the near-infrared, with a slower decay rate — corresponding to the \textit{red kilonova}. These observations provided the first unambiguous association between a kilonova transient powered by r-process nucleosynthesis and a BNS merger. The early blue kilonova is attributed to the outermost ejecta layers expanding at velocities of $\sim0.25c$. The slower inner layers, moving at $\sim0.1c$ and enriched with lanthanides and actinides from the accretion torus surrounding the remnant, are responsible for the red component. AT~2017gfo is associated with the production of heavy elements with mass numbers up to $A \gtrsim 140$\cite{metzger2020kilonovae,radice_GW}.

\subsection{Afterglow: Post-prompt Emission — X-rays to Radio}

Weeks to months following the merger, EM counterparts in the X-ray, radio, and optical bands were detected. Unusually, the emission exhibited a sustained, slow brightening across all wavelengths, peaking approximately 150 days after the merger. This behavior is consistent with an off-axis relativistic jet that gradually decelerates due to interactions with the interstellar medium (ISM), thereby enhancing its visibility along the line of sight. Shock waves generated by the jet–ISM interaction accelerate particles, which subsequently emit synchrotron radiation, accounting for the observed afterglow. Remarkably, the afterglow remains detectable in X-rays even seven years post-merger. Spectral and temporal analyses support a single emission process and indicate that the jet maintains mildly relativistic speeds, with a Lorentz factor of approximately 2\cite{lazzati_GW,gw_afterglow_late_detection}.

\subsection{No Neutrinos}

Neutrinos could be produced during the prompt and extended $\gamma$-ray emission phases through interactions between the jet and the material ejected by the merger, and in possible relativistic winds powered by a neutron star remnant as well. A joint search effort involving the ANTARES, IceCube, and Pierre Auger Observatories was conducted to identify GeV–EeV neutrinos directionally coincident with GW~170817, within a time window of $\pm 500$~s, and in an extended period of up to 14 days following the merger. No neutrino candidates were detected within this energy range. Additionally, no MeV neutrinos associated with the burst were observed. This null detection is consistent with expectations for an off-axis short GRB and aligns with the current model of the event\cite{GW_neutrino}.

\section{The Neutrino KM3-230213A: A Unique Event}
\label{sec:KM3230213A}

The most energetic neutrino detected to date is KM3-230213A, observed by KM3NeT\cite{km3net2025observation} on 13 February 2023 and officially published in February 2025. Between the publication date and May 2025, at least 30 preprints discussing this extraordinary event have appeared on \texttt{arXiv}—an average of $\sim10$ per month.

The detection is inferred from a muon track of energy $120^{+110}_{-60}$~PeV traversing the ARCA detector. The reconstructed trajectory corresponds to an arrival direction passing through approximately 300~km water-equivalent of material, making an atmospheric muon origin highly unlikely. The reconstructed neutrino energy is $220^{+570}_{-110}$~PeV, from right ascension $94.3^\circ$ and declination $-7.8^\circ$, corresponding to $(\ell, b) = (216.1^\circ, -11.1^\circ)$ in Galactic coordinates.

This detection is in tension with the null results reported by IceCube and the Pierre Auger Observatory in the same energy range. Given ARCA's effective area and the reconstructed energy, the inferred per-flavor flux of ultra-high-energy (UHE) neutrinos is $5.8^{+10.1}_{-3.7} \times 10^{-8}$~GeV~cm$^{-2}$~s$^{-1}$, a level that should have been detectable by both IceCube and Auger. Assuming an isotropic flux, the KM3NeT Collaboration\cite{km3net_comparing} assessed the consistency of the KM3-230213A detection with upper limits from IceCube and Auger. A joint fit incorporating all three experiments yields a lower flux, $7.5^{+13.1}_{-4.7} \times 10^{-10}$~GeV~cm$^{-2}$~s$^{-1}$~sr$^{-1}$, suggesting the possibility that KM3-230213A is an upward fluctuation.

The tension between experiments is closely linked to the origin of the neutrino. Within standard astrophysical models, neutrinos are produced exclusively via hadronic interactions. In principle, the event could originate from cosmic-ray interactions with background radiation during propagation (cosmogenic neutrino), or from an astrophysical source—Galactic or extragalactic, steady or transient. The tension is mitigated if a transient origin is assumed\cite{zhang2025cosmogenic,li2025clashtitans,neronov2025_blazar}. Specifically, the discrepancy between KM3NeT and IceCube decreases from $3.5\sigma$ and $3.1$–$3.6\sigma$ (for diffuse or cosmogenic scenarios) to $2.9\sigma$ for stationary point sources and $2.0\sigma$ for transient point sources\cite{li2025clashtitans}.

Das et al.\cite{das2025cosmicrayconstraintsfluxultrahighenergy} analyzed three scenarios for the origin of KM3-230213A: a transient point source, diffuse astrophysical emission, and cosmogenic neutrinos from UHECR point sources. Their results indicate that all scenarios face challenges in explaining the observed flux based solely on the KM3NeT event. However, the tension is alleviated when the joint flux from KM3NeT and IceCube is considered. When GRBs are assumed as transient sources, the required energy injection rate is evaluated. While short GRBs and low-luminosity long GRBs fall short of the energy budget, high-luminosity long GRBs can meet the requirement for redshifts $z \gtrsim 0.2$, though such cases would imply a detectable $\gamma$-ray flux. This suggests the possibility of a new class of sources capable of accelerating particles to $\sim10^{18}$~eV, whose $\gamma$-ray emission is heavily absorbed in the source environment. If the neutrino originates from cosmogenic interactions by a point source—such as a highly collimated UHECR jet—the required luminosity points to powerful blazars as potential progenitors. In the case of unresolved diffuse sources, the source population must extend to redshifts $z \gtrsim 1$ to remain consistent with UHECR constraints from Auger.

Identifying a point source depends on establishing a robust temporal and directional correlation between the neutrino and an electromagnetic counterpart. Despite several proposed candidates, no definitive association has been made to date, and the possibility of an undiscovered source remains open\cite{li2025clashtitans,muzio2025emergenceneutrinoflux}. The absence of a counterpart has also motivated consideration of Beyond Standard Model (BSM) explanations. The following subsections review the various hypotheses proposed for the origin of this event.

\subsection{Astrophysical Hypotheses for the Origin}

\textbf{Galactic sources.}

A Galactic origin for KM3-230213A is considered highly unlikely\cite{adriani2025_galactic}. Producing a UHE neutrino in the Galactic environment requires both a highly efficient cosmic-ray accelerator, capable of reaching $\gtrsim100$~PeV, and a sufficiently dense gas target for hadronic interactions. Such conditions would also generate a detectable $\gamma$-ray counterpart. Adriani et al.\cite{adriani2025_galactic} investigated potential Galactic sources within a $3^\circ$ containment region ($99\%$ CL) around the reconstructed direction and identified the Monoceros R2 molecular cloud as a potential target. However, no plausible accelerators were found nearby—such as supernova remnants, stellar clusters, X-ray binaries, microquasars, or pulsar wind nebulae. This lack of coincidence, coupled with the general difficulty of achieving $\gtrsim100$~PeV energies in Galactic sources, and strong constraints from HAWC and LHAASO $\gamma$-ray observations, disfavor a Galactic origin.

\textbf{Cosmogenic neutrino.}

Despite the $\sim3\sigma$ tension between KM3NeT and IceCube under isotropic assumptions, a cosmogenic origin remains viable. Cosmogenic neutrinos are produced by UHECR interactions with cosmic background photons during propagation. The resulting flux is constrained by UHECR observations from the Pierre Auger and Telescope Array Observatories.

A detailed study by the KM3NeT Collaboration\cite{thekm3netcollaboration2025cosmogenic}, based on a combined fit to Auger's spectrum and composition data, concluded that a neutrino flux matching KM3-230213A is achievable if distant sources ($z_{\rm max} = 6$) with strong positive source evolution and a non-negligible proton fraction at the highest energies are assumed.

Muzio et al.\cite{muzio2025emergenceneutrinoflux} jointly fit UHECR and neutrino data from Auger, IceCube, and KM3NeT. They found that KM3-230213A could be explained by UHECR interactions, but the conclusions depend on the assumed neutrino energy. If $E \approx 100$~PeV, a peak at $\sim30$~PeV is expected due to interactions at the source, causing a deviation from the IceCube spectrum below 1~PeV. If $E \approx 1$~EeV, an additional UHECR source population injecting protons above $\sim10$~EeV is required. This scenario implies a lighter UHECR composition at the highest energies, potentially supported by Auger's $X_{\rm max}$ distributions. These hypotheses can be tested with $\sim15$ years of IceCube-Gen2 data.

\textbf{Extragalactic, transient sources.}

As discussed, transient extragalactic sources reduce the tension between KM3NeT and other experiments. A major challenge lies in the directional uncertainty ($3^\circ$ at 99\% CL). The hadronic origin implies co-production of $\gamma$ rays, motivating multi-wavelength searches.

Crnogor{\v c}evi{\'c} et al.\cite{crnogorcevic2025_cascade} conducted a spatial and temporal scan of \textit{Fermi}-LAT data, but found no significant coincident $\gamma$-ray excess. The absence of such a signal may be due to electromagnetic cascades within the source or intergalactic medium\cite{neronov2025_blazar,dzhatdoev2025_blazar,Fang_2025_cascade}. Cascades can cause time delays and flux dilution. For instance, if the IGMF has strength $B = 0.03$~nG and coherence length $\lambda_B = 1$~Mpc, $\gamma$-ray delays of hours to days are expected; for $B \gtrsim 0.3$~nG, delays can exceed 10 years\cite{Fang_2025_cascade}. Crnogor{\v c}evi{\'c} et al. constrain $B \gtrsim 0.01$~nG for sources at $z < 1$.

Alternatively, the source could be opaque to VHE $\gamma$ rays. A dense radio background would be required for absorption at $\sim140$~MHz, potentially producing detectable low-frequency radio emission\cite{crnogorcevic2025_cascade}. Filipovi{\'c} et al.\cite{filipovic2025_radio} searched ASKAP and VLASS catalogs within $1.5^\circ$, identifying 185 sources, 10 of which are blazar candidates. Three notable objects include:

\begin{itemize}
    \item \textit{Phaedra} (UGCA~127), a nearby galaxy ($8.3 \pm 0.6$~Mpc) with possible AGN-driven starburst regions;
    \item \textit{Hebe} (WISEA~J061715.89-075455.4), a double radio galaxy at $550 \pm 38$~Mpc;
    \item \textit{Narcissus} (EMU~J062248-072246), a variable broadband-spectrum source.
\end{itemize}

If the neutrino is associated with UHECR acceleration, proton energies of at least $\sim1.5$~EeV are required\cite{km3netcollaboration2025_blazars}. The absence of PeV neutrinos during flaring episodes suggests hard proton spectra or production within dense photon fields such as AGN tori\cite{neronov2025_blazar}. Nerov et al.\cite{neronov2025_blazar} find that transient events lasting $\lesssim 2$~yr and releasing $10^{50}$–$10^{54}$~erg are compatible with the KM3NeT detection and null results from IceCube and Auger.

Blazar activity has been widely studied as a potential origin\cite{km3netcollaboration2025_blazars,dzhatdoev2025_blazar}. The KM3NeT Collaboration identified 17 blazars within the 99\% CL region. Cross-referencing temporal flares yields three prime candidates:

\begin{itemize}
    \item \textit{PKS 0605-085}, showing enhanced $\gamma$-ray and possible optical activity in the preceding year; also among the 50 brightest VLBI radio sources;
    \item \textit{PMN~J0606-0724}, with a significant radio flare just five days after the event; pre-trial $p$-value estimated at 0.26\%;
    \item \textit{MRC~0614-083}, displaying steadily increasing X-ray emission in eROSITA data and a persistent high state in Swift-XRT follow-up.
\end{itemize}

Despite several promising candidates, the large positional uncertainty and lack of a definitive temporal correlation preclude an unambiguous association. AGNs, star-forming galaxies, and new classes of transient sources remain viable possibilities.

\subsection{Beyond Standard Model}

\textbf{Dark Matter (DM) origin.}

Given the challenges in associating KM3-230213A with a known astrophysical event, several models propose a super-heavy DM origin\cite{choi2025_DM,khan2025_DM,borah2025_DM,brdar2025_DM,jiang2025_DM,kohri2025_DM,barman2025_DM,murase2025_DM,narita2025_DM}. These include decay of sterile neutrinos, pNGBs, or inflatons, and predict accompanying signals such as gravitational waves (GWs) from cosmic defects or black hole evaporation. Typical DM masses are $\gtrsim440$~PeV, and decay lifetimes are constrained to $10^{29}$–$10^{30}$~s by neutrino observations.

Kohri et al.\cite{kohri2025_DM} proposed a minimal seesaw extension with a singlet scalar and fermion DM. Barman et al.\cite{barman2025_DM} considered a $U(1)$ extension of the Standard Model, yielding right-handed neutrino DM. Jiang \& Huang\cite{jiang2025_DM} proposed pNGB production via Hawking radiation. Murase et al.\cite{murase2025_DM} modeled inflaton decay producing UHE neutrinos and the Amaterasu event. Khan et al.\cite{khan2025_DM} proposed right-handed neutrino portal decay of Dirac fermion DM. Choi et al.\cite{choi2025_DM} studied sterile neutrinos from primordial black holes.

Brdar \& Chattopadhyay\cite{brdar2025_DM} showed that keV-scale sterile neutrinos could oscillate into active states within $\sim150$~km—consistent with KM3NeT detection and IceCube non-detection, due to differing Earth path lengths.

\textbf{Lorentz Invariance Violation (LIV).}

KM3-230213A has also been used to constrain LIV. Satunin\cite{satunin2025_LIV} placed limits of $\Lambda_1 = 1.1 \times 10^{30}$~GeV and $\Lambda_2 = 1.1 \times 10^{19}$~GeV, assuming extragalactic origin and absence of neutrino splitting. Yang et al.\cite{yang2025_LIV} found $\Lambda_2 > 5 \times 10^{19}$~GeV via statistical fits. These constraints improve upon previous IceCube-based limits.

KM3-230213A also constrains the relative neutrino velocity $\delta = c_\nu^2 - 1$ to $\delta < 4.2 \times 10^{-22}$\cite{km3netcollaboration2025_LIV}, improving upon earlier bounds of $\delta < 5.2 \times 10^{-21}$\cite{icecube_LIV2014}.

Amelino-Camelia et al.\cite{amelinocamelia2025_LIV} explored a potential LIV-induced delay between GRB~090401B and KM3-230213A, separated by 14 years and $1.4^\circ$. They inferred an in-vacuo dispersion scale of $M_{QG} = (3.97$–$9.60) \times 10^{17}$~GeV. Wang et al.\cite{Wang_2025_LIV_GRB} extended this to broader GRB samples, yielding LIV constraints up to $5.6 \times 10^{19}$~GeV.

Cattaneo\cite{cattaneo2025_LIV} analyzed the muon lifetime detected by ARCA to constrain maximal attainable velocities and helicity-dependent LIV parameters, obtaining $|c_\mu - c_e| < 1.5 \times 10^{-21}$ and $\epsilon_L^2 + \epsilon_R^2 < 4.8 \times 10^{-28}$.

\section{Final comments}

Models for the origin of UHECRs in the radio galaxy Centaurus~A (Cen~A) were discussed in Section~\ref{sec:cenA}. The proximity of Cen~A makes it one of the most extensively studied astrophysical sources, with observations spanning the entire electromagnetic spectrum, and it is considered a promising candidate for UHECR astronomy. This section underscored the importance of multimessenger data in testing current UHECR acceleration models, and how the discovery of a new component in the $\gamma$-ray spectrum and its spatial morphology has reshaped our understanding of particle acceleration in this object. However, despite tentative correlations between current UHECR detections and neutral messengers, several caveats must be considered.

UHECRs experience significant delays relative to neutrinos and $\gamma$ rays. Thus, \textit{contemporaneous} detections of neutral secondaries do not necessarily correlate with UHECRs, as source activity may vary over time. For instance, Cen~A has been proposed to have undergone more powerful activity in the past\cite{matthews_backflow}. Consequently, even if UHE neutrinos and $\gamma$ rays are detected \textit{now} from an object such as Cen~A, this does not guarantee that the currently measured UHECRs originate from it. The value of such detections lies in demonstrating that an object—or a specific region within it—can act as a source of UHECRs, thereby potentially unveiling the acceleration mechanism and identifying a class of sources.

To relate TeV-scale secondaries with UHECRs, many studies extrapolate the flux across energy scales. Although this is useful for first-order approximations, such extrapolations may not be reliable. Acceleration mechanisms, source environments, and diffusion coefficients may differ across energy regimes, leading to distinct energy spectra for different messengers\cite{Seo_2023,Mbarek_2021}.

Despite the utility of multimessenger comparisons for investigating the origin of UHECRs, a degree of caution is essential. Kachelrieß et al.\cite{Kachelrieß_2009} outlined several limitations of using neutral messengers to trace UHECR origins: (1) the ambiguity in identifying the hadronic or leptonic origin of $\gamma$ rays, noting that VHE $\gamma$ rays are significantly attenuated both at the source and during extragalactic propagation; (2) the extremely long mean free path of neutrinos, which, when combined with their poor angular resolution (typically $\sim 1^\circ$) and the low expected event rate, makes source identification via neutrinos alone highly challenging, thus requiring complementary information such as timing and directionality from other messengers. Today, gravitational wave observations offer a valuable addition to this multimessenger framework.

An insightful discussion on the \textit{necessary} and \textit{sufficient} conditions for establishing the origin of UHECRs was provided by Batista\cite{batista2025quest}, to which we refer the interested reader.

Section~\ref{sec:GW170817} briefly presented the event GW~170817 from a multimessenger perspective. This event exemplifies the scientific potential of combining different messengers, and how gravitational-wave observations are crucial for probing astrophysical processes and compact object dynamics. One of the key elements that enabled the success of this observation was the coordinated alert system triggered by the GW and $\gamma$-ray detections. Despite the significant advances achieved, many questions remain unresolved, highlighting the need for continued instrumental development and an interconnected observational network for transient phenomena.

In Section~\ref{sec:KM3230213A}, the KM3~230213A event was discussed, along with the scientific momentum it generated. On the observational side, the event raised tensions due to the discrepancy between KM3NeT’s detection and the null results from other observatories, compounded by its large angular uncertainty and the absence of a coincident electromagnetic transient. On the theoretical side, the effort to interpret the nature of the event and the lack of an accompanying EM signal led to hypotheses involving known astrophysical classes, previously unidentified sources, and even new physics scenarios. One of the main disappointments associated with this event was the absence of X-ray data at the time of the neutrino detection. Such data could have supported a compelling hypothesis involving MRC~0614-083 and potentially resolved the ambiguity. This highlights the necessity of comprehensive, multi-wavelength observational coverage for interpreting unexpected events.

In light of the increasing volume of publications following each significant discovery, how can the scientific community effectively organize this influx of information and transform it into coherent knowledge? How can we prevent major discoveries from being diluted into "noise" within journals and preprint repositories? Such concerns have been increasingly discussed within the community\cite{pnas_science,hanson_strain_science,pan_inflation_science}, and are expected to intensify as the next generation of experiments yields ever-larger datasets.

\newpage
\section*{Acknowledgments}
CdO acknowledges Henrique Malavazzi for continuous discussions  on the growth of scientific publications and the organization of knowledge. This study was financed, in part, by the São Paulo Research Foundation (FAPESP), Brasil. Process Number 2025/03325-5, 2021/01089-1, 2020/15453-4 and 2019/10151-2. VdS is also supported by CNPq. The authors acknowledge the National Laboratory for Scientific Computing (LNCC/MCTI, Brazil) for providing HPC resources of the SDumont supercomputer, which have contributed to the research results reported within this paper. URL: \url{http://sdumont.lncc.br}
%\end{acknowledgments}

\bibliographystyle{ws-ijmpa}
\bibliography{sample}

\end{document}